\documentclass[prl,aps,twocolumn,superscriptaddress]{revtex4}  

\usepackage{graphicx}

\begin{document}

\title{Are fluorinated BN nanotubes $n$-type semiconductors?}

\author{H. J. Xiang}
\affiliation{Hefei National Laboratory for Physical Sciences at
  Microscale, 
  University of Science and Technology of
  China, Hefei, Anhui 230026, People's Republic of China}
\affiliation{USTC Shanghai Institute for Advanced Studies,
  University of Science and Technology of China,
  Shanghai 201315, People's Republic of China}

\author{Jinlong Yang}
\thanks{Corresponding author. E-mail: jlyang@ustc.edu.cn}
\affiliation{Hefei National Laboratory for Physical Sciences at
  Microscale, 
  University of Science and Technology of
  China, Hefei, Anhui 230026, People's Republic of China}
\affiliation{USTC Shanghai Institute for Advanced Studies,
  University of Science and Technology of China,
  Shanghai 201315, People's Republic of China}

\author{J. G. Hou}
\affiliation{Hefei National Laboratory for Physical Sciences at
  Microscale, 
  University of Science and Technology of
  China, Hefei, Anhui 230026, People's Republic of China}
\affiliation{USTC Shanghai Institute for Advanced Studies,
  University of Science and Technology of China,
  Shanghai 201315, People's Republic of China}

\author{Qingshi Zhu}
\affiliation{Hefei National Laboratory for Physical Sciences at
  Microscale, 
  University of Science and Technology of
  China, Hefei, Anhui 230026, People's Republic of China}
\affiliation{USTC Shanghai Institute for Advanced Studies,
  University of Science and Technology of China,
  Shanghai 201315, People's Republic of China}

\begin{abstract}
The structural and electronic properties of fluorine (F)-doped 
BN nanotubes (BNNTs) are studied using density functional
methods. Our results indicate that F atoms prefer to substitute
N atoms, resulting in substantial changes of BN layers.
However, F substitutional doping results in no shallow impurity
states. The adsorption of F atoms on B sites is more stable than
that on N sites. BNNTs with adsorbed F atoms are $p$-type
semiconductors, suggesting the electronic conduction in F-doped
multiwalled BNNTs with large conductivity observed
experimentally might be of $p$-type due to the adsorbed F atoms,
but not $n$-type as supposed before.
\end{abstract}

\maketitle

Pure boron nitride
nanotubes (BNNTs) \cite{BNNT1,BNNT2} are semiconductors regardless of diameter,
chirality, or the number of walls of the tubes.\cite{BNNT1} This
contrasts markedly with the heterogeneity of electronic properties
of carbon nanotubes, and also makes BNNTs particularly useful for potential 
nanoelectronic applications.
From the applications viewpoint, obtaining n-type and/or p-type
semiconducting BNNTs is very important for the use in the nanoscale
electronic device, such as
the $p$-$n$ junction, $n$-type or $p$-type nanoscale field-effect
transistors.
In this context, the uniformly doped BNNTs
obtained through chemical modification would be a
prospective object for tailoring the electronic properties.
Theoretical studies indicated that H adsorption on B atom or N atom of
BNNTs will introduce donor or acceptor states respectively.\cite{BNH1}
Previous experimental studies on functionalization of BNNTs have
mainly focused on the B-N-C system through the sophisticate control of
C content.\cite{BNC} 
Very recently, Tang {\it et al.} \cite{BNF} have successfully obtained
F functionalized BNNTs with highly curled
tubular BN sheets, whose resistivity is smaller by
three orders of magnitude than that of the undoped BNNTs.
And based on the sole fact that F
has excessive valence electrons compared to B and N, 
the F-doped BNNTs were supposed to be n-type semiconductors.\cite{BNF}
However, the behavior of dopants in semiconductors is often far from
trivial.\cite{doping} 
Then does F-doping really induce $n$-type conduction in
BNNTs?

In this Letter, we conduct a first-principles study on F-doped BNNTs
to address this issue. 
Both adsorption and substitution of F on BNNTs
are examined. Surprisingly, neither adsorption or substitution of F on
BNNTs results in $n$-type semiconductors, contrasting sharply to
previous supposition. 
Our calculations are performed using the
local density approximation\cite{LDA} in the density
functional theory as implemented in VASP \cite{vasp}. 
The electron-ion interaction is
described by ultrasoft pseudopotentials
\cite{ultrasoft} and the cut-off energy is set to $347.9$ eV.

\begin{figure}[!hbp]
  \includegraphics[width=7.5cm]{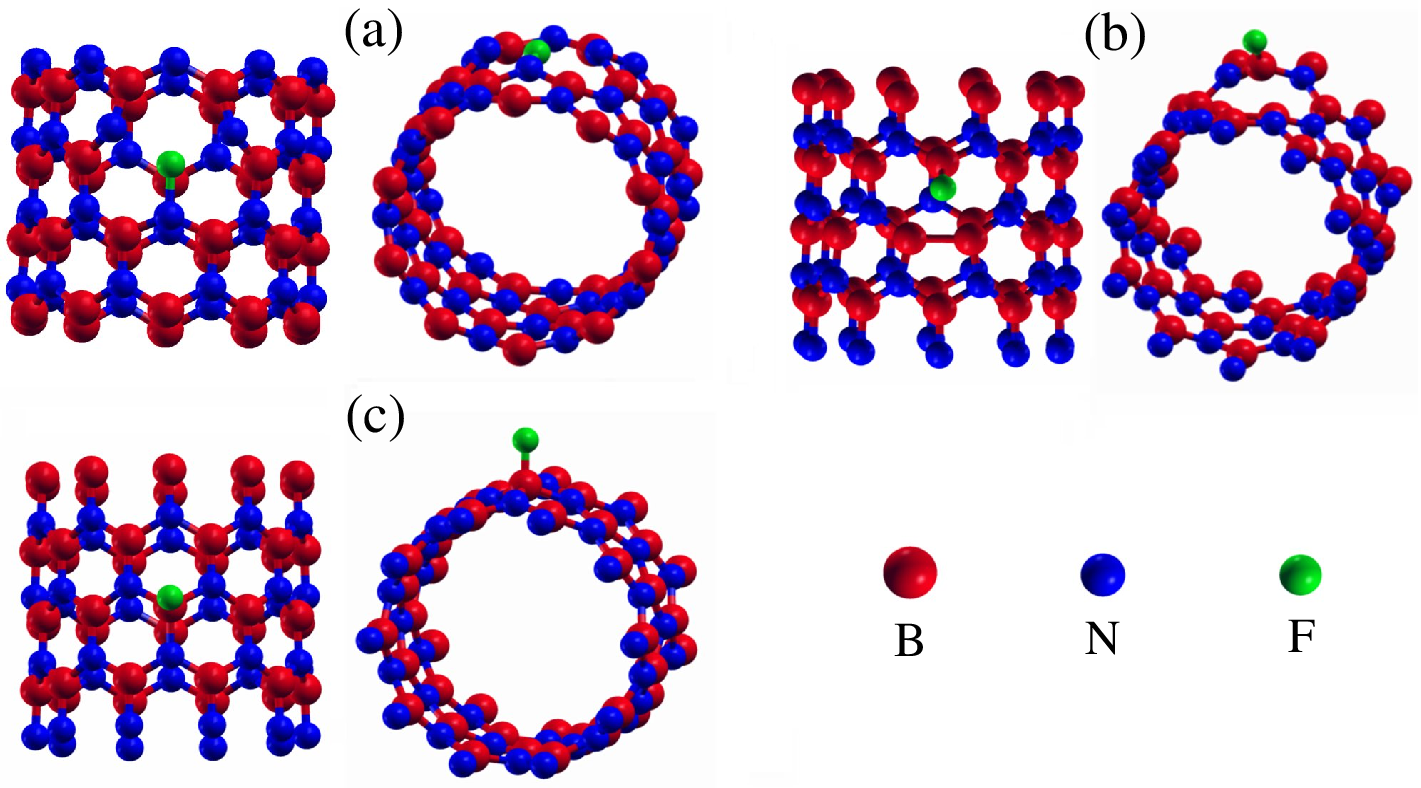}
  \caption{(Color online) Three different configurations of F-doped
    BN(10,0) nanotubes: 
    (a) one B
    substituted by one F atom ($S_F^B$), (b) one N substituted by one F
    atom ($S_F^N$), and (c) one F adsorbed on a B atom ($A_F^B$).
    Both top view and side view are shown.}
  \label{fig1}
\end{figure}

We mainly focus on a typical zig-zag BNNT, i.e., BN(10,0) nanotube.
Test calculations indicate that the doping behavior for BN(5,5)
NT and BN two dimensional sheet (a model for nanotube with
radius $R=\infty$) is qualitatively 
the same as that for BN(10,0) nanotube.
Our supercell for BN(10,0) nanotube contains 
80 atoms.
First from the optimized structure of the pure BN(10,0) nanotube, we
substitute one B or N with one F atom (abbreviated as $S_F^B$ and $S_F^N$
respectively). Both relaxed structures are 
shown in Fig.~\ref{fig1}(a) and (b) respectively. In both cases, F
substitution induces significant deformation of the BN layer.  
Since F is usually in $-1$ valence state, F is
bonded with only one adjacent B or N atom along the axis direction.
A remarkable difference between the two structures is that F atom
doesn't lie in the BN layer and the other two B atoms with dangling bonds
come close to each other to form a homoelemental bond in $S_F^N$.
Another difference is that the cross section of the tube
in $S_F^N$ is no longer a perfect circle 
as the B atom bonded with F comes outer from the BN plane.
To see which substitution is more favorable, we compare the defect
formation energy in these two cases.
The formation energy is defined as:
  $E_{form}=E_{tot}-n_B \mu _B - n_N \mu _N -n_F \mu _F,$
where $E_{tot}$ is the calculated total energy of the system, $n_B$,
$n_N$, and $n_F$ are the number of B, N, and F atoms respectively, 
and $\mu$ is the chemical potential. $\mu _B$ and
$\mu _N$ depend on the growth condition: In the N-rich environment,
$\mu _N$ ($-8.629$ eV) is obtained from nitrogen in the gas phase,
while a metallic $\alpha$-B phase is used as the reservoir for the
B-rich environment ($-7.491$ eV for $\mu _B$). 
In both cases, $\mu _N$
and $\mu_B$ are linked by the thermodynamic constraint:
  $\mu _B+\mu _N=\mu _{BN}^{tube},$
where $\mu_{BN} ^{tube}$ is the chemical potential per BN
pair in the pristine BN nanotube.
Since both systems has one F atom, the value of $\mu
_F$ makes no difference to the difference of the formation energy.
So, we don't specify $\mu _F$ explicitly. The calculated formation
energies are shown in Table~\ref{table1}. 
The formation energy difference
between $S_F^B$ and $S_F^N$ is $9.778$ ($3.707$) eV in the B (N) rich
environment, suggesting F strongly
prefers to substitute N in BNNTs. This is not surprising since the large
electronegativity of F will result in stronger F-B bond. The
homoelemental bond in $S_F^N$ partially saturates the dangling bonds of
B atoms, also contributing the stability of $S_F^N$.

\begin{figure}[!hbp]
  \includegraphics[width=7.5cm]{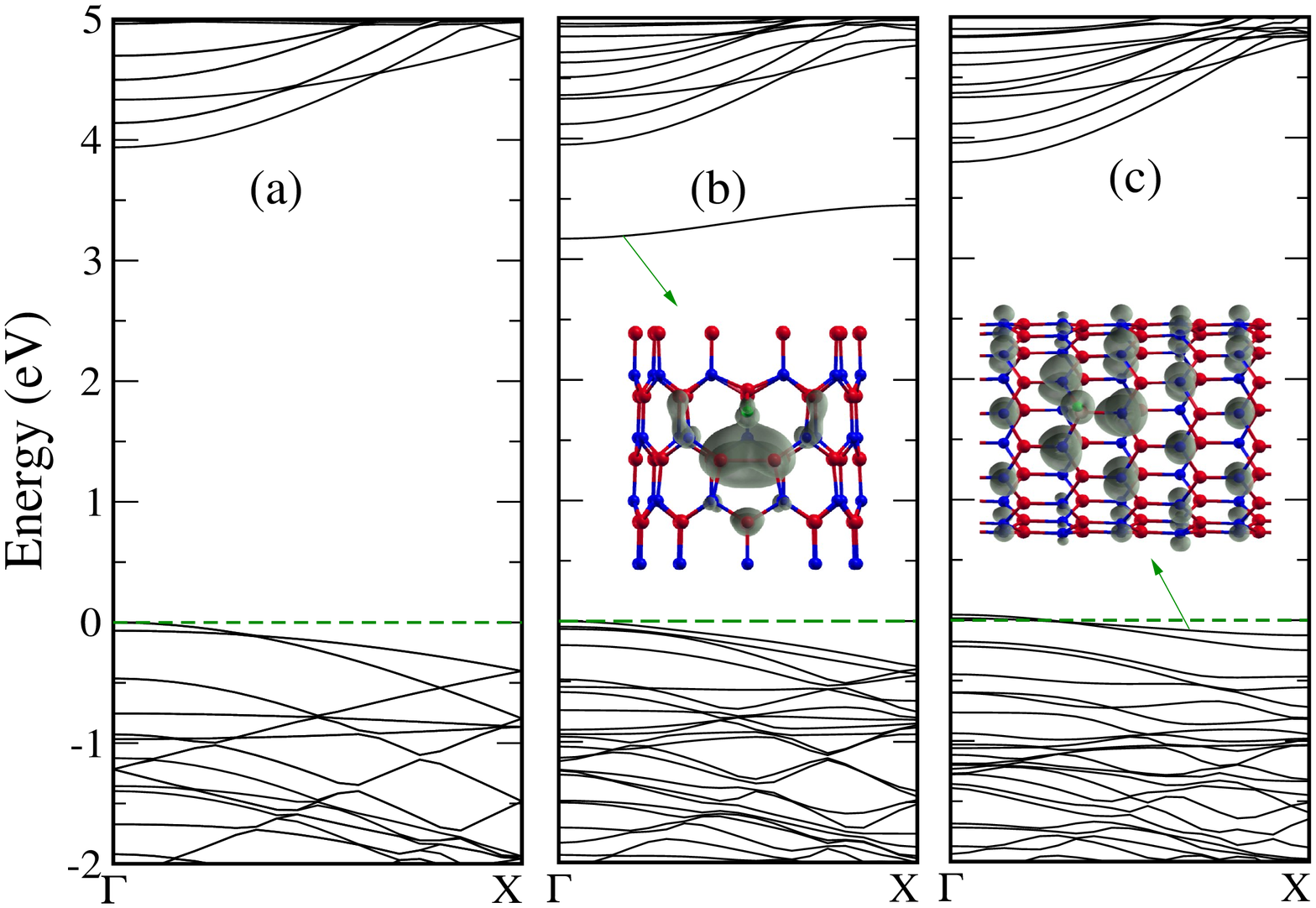}
  \caption{(Color online) Band structures for (a) pristine
    BN(10,0) nanotube, (b) BN(10,0) nanotube with one N
    substituted by one F atom, and (c) BN(10,0) nanotube with
    one F atom adsorded on one B atom. The local DOS for the states
    induced by dopants are shown in the insets.
    The valence top or the Fermi level is taken as zero-enegy point in
    (a) and (b) or (c) respectively.}
  \label{fig2}
\end{figure}

The band structure of $S_F^N$ is shown in
Fig.~\ref{fig2}. For better comparison, the band structure of
the pristine BN(10,0) nanotube with the same supercell is also plotted
in Fig.~\ref{fig2}. 
We can see both pristine and F-doped BN(10,0) nanotubes are
semiconducting with a band gap of $3.9$ and $3.2$ eV respectively.
Density of states (DOS) analysis shows that the bands related to F 2p states
mainly lie in $-4$ to $-8$ eV with respect to the valence top.
F substitutional doping induces an unoccupied localized state in the
gap, which is mainly contributed by the two B atoms connected with the
homoelemental bond as shown in the inset of Fig.~\ref{fig2}(b).
However, this state is neither a shallow acceptor state
nor a shallow donor state. So we don't expect such F substitutional
doping will result in large increase of electrical conductivity.

To elucidate the origin of the drastic increase in the conductivity
of F-doped BNNTs, the 
adsorption of F on BN(10,0) nanotube is examined.
We find that one F atom prefers to adsorb on the B site (labeled as $A_F^B$,
as shown in Fig.~\ref{fig1}(c)), 
since the adsorption of F on N site is unstable and the F atom will 
eventually adsorb on the B site.
The electronic band structure for $A_F^B$ is
shown in Fig.~\ref{fig2}(c). Clearly, the system displays a degenerated
$p$-type semiconducting behavior. There are two bands with similar
character crossing the Fermi level. Both states are rather delocalized
as can be seen from the local DOS of one of the bands (shown in the inset
of Fig.~\ref{fig2}(c)). 
So only 1.25\% F adsorption on BNNTs
will lead to drastic increase in conductivity.

\begin{figure}[!hbp]
  \includegraphics[width=7.5cm]{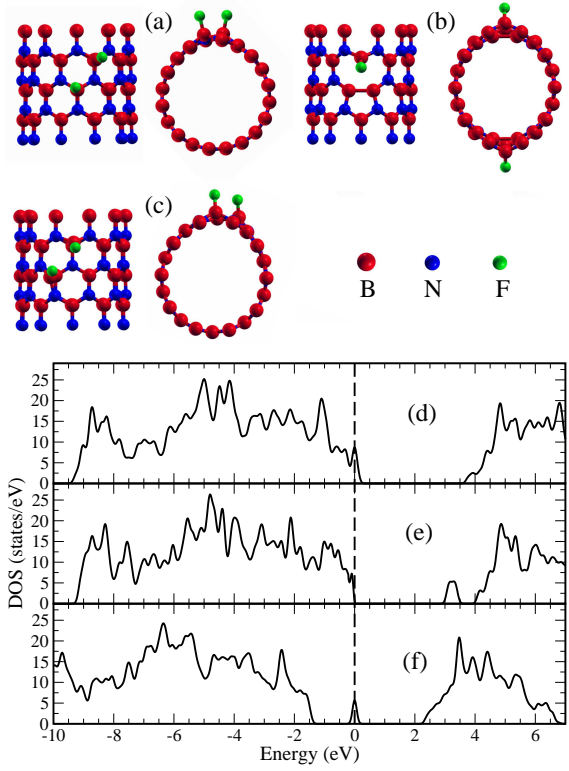}
  \caption{(Color online) Three different configurations BN(10,0) nanotubes doped
    with two F atoms:
    (a) two F atoms adsorb on two B sites,
    (b) two substitutional F atoms, and
    (c) one F atom adsorbed on one surrounding B atoms around the
    substituted N site.
    Both top view and side view are shown.
    Total DOS for these three structures are shown in (d), (e), and
    (f) respectively.
    The Fermi level (or valence top) is taken as zero-enegy point in
    (d) and (f) (or (e)).}
  \label{fig3}
\end{figure}

Besides the doping behavior of one F atom, we also study the BN(10,0)
nanotube doped with two F atoms. First, adsorption of two F atoms is
examined. We find that the configuration (as shown in Fig.~\ref{fig3}(a))
where both F atoms adsorb on two B sites is favourable over the one
where the second impurity atom adsorbs on the adjacent N site. This
differs fundamentally the adsorption of H atoms on BNNTs.\cite{BNH2}
Electronic structure calculations on this configuration
indicate that it is a degenerated $p$-type semiconductor with larger
DOS in the Fermi level than $A_F^B$, as shown in Fig.~\ref{fig3}(d).
Then BN(10,0) nanotube with two substitutional F atoms as shown in
Fig.~\ref{fig3}(b) is studied. The DOS of this configuration is shown in
Fig.~\ref{fig3}(e), indicating that BN(10,0) nanotube with two
substitutional F atoms has similar semiconducting properties as $S_F^N$.
Finally, BN(10,0) nanotubes with one substitutional F atom and one
adsorbed F atom are investigated. 
The  most stable configuation is the
sturucture shown in Fig.~\ref{fig3}(c), where the
excess F atom is bonded with one 
of the two B atoms which form the homoelemental B-B bond in $S_F^N$.
The DOS of this configuration is shown in Fig.~\ref{fig3}(f),
indicating that there is a half occupied localized state
due to the unsaturated B dangling bond at approximately 1.4 eV above
the valence top. 
Thus we don't expect this configration will have high electric
conductivity since there is no shallow impurity related states. 
Besides the most stable configuation, we also consider other
metastable configuations where one more F atom
adsorbs on other B sites in $S_F^N$. All these metastable
structures are found to be p-type
semiconductors as $A_F^B$.

Based on our calculations, we suggest that the experimentally observed
three orders decrease in resistivity of F-doped BNNTs might be due to
F adsorption instead of F substitution. One may concern the stability
of the adsorption of F atoms on BNNTs since experimental studies
indicated that the doped fluorine escapes easily when the tube is
exposed to air or under standard beam irradiation in the electron
microscope.\cite{BNF}
Our results indicate that the adsorption of F on BNNTs is as easy as
the substitution of F for N in BNNTs in the B rich environment, and
the formation of $A_F^B$ is much easier than that of  $S_F^N$ in the N rich
environment, as can be seen from their formation energies.
Our nudged elastic band (NEB) \cite{NEB} calculations indeed show that 
the F dissociation in single-walled BNNTs with adsorption
of F atoms is easier than that in $S_F^N$ (with dissociation energy
$3.44$ eV and $5.86$ eV for $A_F^B$ and $S_F^N$ respectively).
However, Tang {\it et al.} synthesized F-doped BNNTs through
the introduction of F atoms at the stage of multiwalled nanotube
growth \cite{BNF} instead of doping F after the stage of BNNTs
growth. We suppose 
that F adsorption occur in the stage of BNNTs 
growth, the F adsorbed on the inner walls of the multiwalled BNNTs
should be stable. 
However, our calculations don't exclude the possibility of F
substitution: F substitutional doping might also occur along with F
adsorption, especially in the B rich environment. In fact, F
substitutional doping destroys the six-numbered BN atomic ring, which
is in accord with experimental findings.\cite{BNF}  

To summarize, first-principles calculations have been performed to
study the structural and electronic properties of fluorinated BNNTs.
F atoms prefer to substitute N atoms, resulting in substantial
structural changes of the BN layers but little change in
conductivity. The adsorption of F on B 
sites is more favourable than 
that on N sites due to the large electronegativity of F.
BNNTs with adsorbed F atoms are $p$-type semiconductors with
enhanced conductivity than the pristine nanotubes. Our results clearly
demonstrate that F-doped multiwalled BNNTs synthesized experimentally
are most likely $p$-type semiconductors. The present study also
suggests that care should be taken when predicting from intuition
doped semiconductors' electronic properties.
Further experimental studies on F-doped BNNTs are awaited to
confirm our results.

This work is partially supported by the National
Project for the Development of Key Fundamental Sciences in China
(G1999075305), by the National Natural Science Foundation of China
(50121202, 10474087), by the USTC-HP HPC project, by the EDF of
USTC-SIAS, and by the SCCAS.

\clearpage

\begin{table}[!hbp]
  \caption{Defect formation energy (E$_f$) for various doped BNNTs (We regard
    the BN sheet as a nanotube with radius $R=\infty$).
    For each of the three nanotubes, the first and second
    lines are the formation energies in the B rich or N rich
    environment respectively. Please refer to the text for the
    definition of $S_F^B$, $S_F^N$, and $A_F^B$.
    Energy is in eV.} 
  \begin{tabular}{ccccc}
    \hline
    \hline
           & & E$_f$($S_F^B$)+$\mu_F$ &E$_f$($S_F^N$)+$\mu_F$&
    E$_f$($A_F^B$)+$\mu_F$ \\
    \hline
    BN(10,0)&B rich& 5.887 &$-$3.891  &$-$3.846  \\
            &N rich& 2.851 &$-$0.856  &$-$3.846  \\
    \hline
    BN(5,5) &B rich& 5.851 &$-$3.907  &$-$3.995  \\
            &N rich& 2.868 &$-$0.923  &$-$3.995  \\
    \hline
    BN sheet&B rich& 7.196 &$-$2.453  &$-$3.188  \\
            &N rich& 3.993 &  0.750   &$-$3.188  \\
    \hline
    \hline
  \end{tabular}
  \label{table1}
\end{table}

\clearpage

\end{document}